\documentclass[sn-mathphys-ay]{sn-jnl}

\usepackage{graphicx}%
\usepackage{multirow}%
\usepackage{amsmath,amssymb,amsfonts}%
\usepackage{amsthm}%
\usepackage{mathrsfs}%
\usepackage[title]{appendix}%
\usepackage{xcolor}%
\usepackage{textcomp}%
\usepackage{manyfoot}%
\usepackage{booktabs}%
\usepackage{algorithm}%
\usepackage{algorithmicx}%
\usepackage{algpseudocode}%
\usepackage{listings}%
\usepackage{moreverb,url}

\begin{document}

\title[FDA on Wearable Sensor Data:A Systematic Review]{Functional Data Analysis on Wearable Sensor Data: A Systematic Review}

\author*[1]{\fnm{Nihan} \sur{Acar-Denizli}}\email{nihan.acar.denizli@upc.edu}

\author[1]{\fnm{Pedro} \sur{Delicado}}\email{pedro.delicado@upc.edu}

\affil*[1]{\orgdiv{Department of Statistics \& Operational Research}, \orgname{Universitat Polit\`ecnica de Catalunya}, \city{Barcelona}, \country{Spain}} 

\abstract{Wearable devices and sensors have recently become a popular way to collect data, especially in the health sciences. The use of sensors allows patients to be monitored over a period of time with a high observation frequency. 
Due to the continuous-on-time structure of the data, novel statistical methods are recommended for the analysis of sensor data. One of the popular approaches in the analysis of wearable sensor data is functional data analysis.
The main objective of this paper is to review functional data analysis methods applied to wearable device data according to the type of sensor. In addition, we introduce several freely available software packages and open databases of wearable device data to facilitate access to sensor data in different fields.}




\maketitle

\section{Introduction}
\label{sec1}

Digitization and recent technological advances are making way for new methods of data collection.
From environmental studies to health, remote sensor data collection has become an increasingly popular way to obtain data.
Wearable technology permits the monitoring of people's health and physical activity status and facilitates the diagnosis of many important diseases such as Type II diabetes.
Especially with the integration of sensors to smart devices, even in our daily life, we can record many physiological characteristics such as stress level,
heartbeat and daily number of steps by using our smartphones.

Different types of sensors are used in wearable devices, depending on the characteristic that is intended to be measured.
{\em Pedometers} are used to count steps.
{\em Accelerometers}, which are able to monitor any kind of movement, have wider range of use, from counting steps to monitoring heart beats.
{\em Glucometers} are used to monitor blood glucose levels while {\em heart rate sensors} control heart rate beats.
There are also wearable {\em tensiometers} that measures hourly systolic blood pressure, wearable {\em spirometers} that monitor respiratory activities of the people, and {\em electrodermal activity sensors} used to measure changes in the electrical properties of the skin, mainly associated with changes in the emotional state or with metabolic activity.

Common analytical approaches applied to the features extracted from wearable sensors are standard statistical methods that consider them as either multivariate or longitudinal data.
Recently, functional data analysis (FDA) approaches are becoming more popular as an alternative to classical multivariate data approaches and the application of FDA on wearable sensor data is getting more common. Two review papers on physical activity data obtained from accelerometers mentioned some of the FDA approaches and their advantages.
The first of them \citep{Zhang2019review} reviewed both the longitudinal and FDA methods that were used to assess the impact of energy expenditure on body mass index among elementary school-aged children.
The paper introduced the general concepts of multivariate and multilevel functional data models.In the other paper \citep{Migueles2022} the statistical methods used in analyzing accelerometer data were reviewed and the advantages of FDA were discussed. The FDA approach, with fewer assumptions than the other classical analytical methods, was mentioned as an alternative way of analyzing acceleration functions and revealing the association between physical behaviour and health outcomes. In a different context (remote sensor data), the connection between multivariate and functional methods has been recently reviewed by \citet{Li2022}, considering the developments in multi-level functional data analysis, high-dimensional functional regression, and dependent functional data analysis.

The present study has two main objectives.
The first goal is to conduct a systematic literature review on functional data approaches to wearable device data analysis, according to sensor type and application area.
The second goal is to provide a list of open databases of wearable device data, as well as free software tools for analyzing them.
Even though there exist two review papers on statistical methods for accelerometer data (as we have mentioned before: \citeauthor{Zhang2019review}, \citeyear{Zhang2019review}, and
\citeauthor{Migueles2022}, \citeyear{Migueles2022})
and one for glucometer and electrocardiography (ECG) data simultaneously recorded \citep{charamba2023},
we consider it is worthy to perform such a systematic literature revision
as the one we make here for several reasons:
(i) we focus on contributions on FDA with a total of 59 revised papers, while in the reviews of
\citet{Zhang2019review}, \citet{Migueles2022}, and \citet{charamba2023}
there were, respectively, only 9, 4, and 1 references related to FDA;
(ii) at our knowledge, this is the first time that all kind of wearable devices are covered simultaneously in a review paper;
(iii) we offer a detailed explanation of each publication including
information on the used statistical methodology, dataset and wearable technology.

In Section \ref{sec:FDA} we introduce the most common tools in FDA (functional principal component analysis and functional linear models).
Section \ref{sec:BibSearch} describes the literature review process done on Web of Science
\citep[WoS,][]{WoS} and Scopus \citep{Scopus}, indicating keywords used for the bibliographic search.
The results of the literature review are summarized in three Sections corresponding to different types of sensors:
accelerometer (Section \ref{sec:Accel}), glucometer (Section \ref{sec:Gluco}) and other sensors (Section \ref{sec:Other}).
Section \ref{sec:Open} introduces various free accessed software packages and open databases of wearable device data.
Finally, we summarize our conclusions in Section \ref{sec:Concl}.

\section{Functional Data Analysis}\label{sec:FDA}
A random variable $\mathcal{X}$ taking values in a functional space (i.e., a set of functions) is said to be functional.
An observation of $\mathcal{X}$ is called functional data.
A functional data set $\mathcal{X}_1,\ldots,\mathcal{X}_n$ is the
set of $n$ independent observations of $\mathcal{X}$.
It is usual to work with functional data defined in an interval
$T=[a,b]\subseteq \mathbb{R}$ that are square integrable, so they are elements of
$$
L^2(T)=\{f:T\rightarrow \mathbb{R}, \mbox{ such that } \int_T f(t)^2 dt < \infty\}.
$$
The functional space $L^2(T)$ with the inner product
$\langle f,g \rangle = \int_T f(t)g(t) dt$
is a separable Hilbert space, an infinite-dimensional generalization of the usual Euclidean spaces $\mathbb{R}^p$.
In practice, it is not possible to observe a functional data $\mathcal{X}_i(t)$ for all $t\in T$. Instead, it is recorded at a finite number of points $t_1,\ldots, t_m$ in $T$, which are usually densely spread over the whole interval $T$ and are common to all the observed functional data.
The books of \citet{Ramsay2005} and \citet{KokRei:2017} offer excellent introductions to FDA.
A functional data $\mathcal{X}(t)$ can be approximated by a linear combination of $K$ known basis functions $\phi_k(t)$,
\begin{equation}
    \mathcal{X}(t)\approx \sum_{k=1}^K c_k \phi_k(t).
\end{equation}
The most common types of basis functions used in FDA are B-spline basis functions.
Other alternative methods such as wavelet basis functions, kernel smoothing, functional principal components, and functional partial least squares basis functions could be used as well to represent functional data.
Among them, functional principal components and partial least squares basis functions are obtained from dimensionality reduction of functional data.
In Section \ref{sec:FPCA} we introduce Functional Principal Components Analysis (FPCA).
Functional Partial Least Squares (FPLS) is similar to FPCA, but is specifically designed for the case where a scalar response is observed in addition to the functional data.
In Section \ref{sec:FLRM} we introduce different types of
functional regression models which have been used to analyze wearable device data.

\subsection{Functional Principal Component Analysis} \label{sec:FPCA}
FPCA helps to represent a functional data in terms of a combination of orthonormal variables that are obtained by maximizing the variance of the component scores. Therefore, they reveal the most important modes of variation in the variables \citep{Ramsay2005}.

The FPCA model of a functional data set is based on the Karhunen-Lo\`{e}ve (KL) expansion which is given as follows,
\begin{equation}
\label{eq:FPCA}
 \mathcal{X}(t)=\mu(t)+ \sum_{k=1}^{\infty} \xi_k \psi_k(t) + \epsilon(t),
\end{equation}
where $\mu(t)$ is the mean function, $\psi_k(t)$ is the eigenfunction associated with the $k$-th eigenvalue $\lambda_k$ obtained from spectral decomposition of covariance operator of $\mathcal{X}$, $\xi_k$ are FPCA scores and $\epsilon(t)$ is a random functional noise 
\citep[see, for instance, ][]{Horvath2012}.
In practice the expression of the summation term in equation (\ref{eq:FPCA}) is truncated to the first $K$ terms:
\begin{equation}
\label{eq:FPCAtrunc}
\mathcal{X}(t) \approx \mu(t) + \sum_{k=1}^K \xi_k \psi_k(t).
\end{equation}

This model was extended to Generalized Functional Principal Components Analysis (GFPCA) to model non-Gaussian functional data by \citet{HallMullerYao:2008}
assuming a known link function $g(\cdot)$. The GFPCA model is defined by
\begin{equation}
\label{eq:GFPCA}
\mathbb E[\mathcal{X}(t)\mid \xi_1\dots \xi_K] = g \Bigl(\mu(t)+\sum_{k=1}^K \xi_k \psi_k(t) \Bigr).
\end{equation}

Later, two-level FPCA was proposed by \citet{di2009multilevel}, with the name of {\em multilevel FPCA}, that allows modeling two repeated observations per subject.
\citet{greven2010longitudinal} extended this proposal to longitudinal FPCA model to cover the case that there would be at least three repeated functional observations per subject \citep{greven2010longitudinal}.
The underlying ideas of these models are as follows. Assume that $n$ patients have been observed at multiple visits ($m_i$ times for patient $i$) and that a function has been observed at each visit.
The two-level functional mixed model is stated as follows:
\begin{equation}
\label{eq:twolevelmixed}
    \mathcal{X}_{ij}(t)= \mu(t) +  Z_i(t) + U_{ij}(t),
    j=1,\ldots, m_i, i=1,\ldots, n,
\end{equation}
where $\mu(t)$ is the overall mean function,
$Z_i(t)$ is the random subject-specific shift from the overall mean function,
and $U_{ij}(t)$ is the residual subject- and visit-specific deviation.
\citet{di2009multilevel} proposed to replace $Z_i(t)$ and $U_{ij}(t)$ in (\ref{eq:twolevelmixed}) by their corresponding KL expansions, conveniently truncated, giving rise to the two-level FPCA model:
\begin{equation}
\label{eq:twolevelFPCA}
    \mathcal{X}_{ij}(t)= \mu(t) +
    \sum_{k=1}^K \xi_{ik}\phi_k^{(1)}(t) +
    \sum_{l=1}^L \zeta_{ij}\phi_l^{(2)}(t),
    j=1,\ldots, m_i, i=1,\ldots, n,
\end{equation}
where $\xi_{ik}$ and $\zeta_{ij}$ are respectively level 1 and level 2 FPC scores and $\phi_k^{(1)}$ and $\phi_l^{(2)}$ are respectively level 1 and level 2 eigenfunctions.
In order to obtain the eigenfunctions in model (\ref{eq:twolevelFPCA}) it is required to estimate the covariance functions
\[
C_Z(s,t)=\mbox{Cov}(Z_i(t),Z_i(s)),\,
C_U(s,t)=\mbox{Cov}(U_{ij}(t),U_{ij}(s)).
\]
Observe that the following relations are verified:
\begin{align*}
\mbox{Cov}(X_{ij}(t),X_{ij}(s)) &= C_Z(s,t) + C_U(s,t),\, \\
\mbox{Cov}(X_{ij}(t),X_{ik}(s)) &= C_Z(s,t) \mbox{ for } j\ne k.
\end{align*}
The left-hand side terms can be estimated by the method of moments
\citep[see details in][]{di2009multilevel}
and $C_U(s,t)$ is estimated as their difference.
The two-level FPCA model admits a generalized version similar to (\ref{eq:GFPCA}).

\citet{shou2015structured} generalize the multilevel and longitudional FPC models under the name of Structured Functional Principal Components Analysis (SFPCA) and generalize FPCA models to multiway nested and crossed design which as well includes an interaction term.

\subsection{Functional Linear Regression Models} \label{sec:FLRM}
A functional linear regression model (FLRM) is an extension of the multiple linear regression model,
\begin{equation}
\label{lm}
Y=\mathbf{x}^{\text{\tiny T}} \boldsymbol{\beta}+\boldsymbol{\epsilon},
\end{equation}
where at least one of the variables included in the model has a functional structure. According to the position of the functional and real random variables in the model, FLMs are divided into three main groups:
the Scalar on Function Regression (SoFR) model (scalar response, functional predictor),
the Function on Scalar Regression (FoSR) model
(functional response, scalar predictors),
and
the Function on Function Regression (FoFR) model (both the response and the explanatory variables are functions).

The SoFR model has the form
\begin{equation}
\label{flm}
Y= \langle \mathcal{X}, \beta \rangle + \epsilon = \int_T \mathcal{X}(s) \beta(s) ds + \epsilon,
\end{equation}
where the predictor $\mathcal{X}$ and the coefficient $\beta$ are square integrable functions taking values at point $t \in T$.

The FoSR model is defined by
\begin{equation}
\label{fosr}
\mathcal{Y}(t)= \beta_0(t) + \sum_{h=1}^p x_{h}\beta_h(t) + \epsilon(t).
\end{equation}
When the scalar predictor is a factor, FoSR becomes a functional ANOVA model known as FANOVA
\citep{Ramsay2005}.

Finally, the FoFR model takes the form
\begin{equation}
\label{fof}
\mathcal{Y}(t)= \int \beta(t,s)\;\mathcal{X}(s)+ \epsilon(t).
\end{equation}

Thereafter, SoFR model was extended to Generalized Functional Linear Regression Models \citep[GFLRM,][]{Muller2005} which enables to model a scalar response that, conditional on a functional predictor, follows a distribution in the exponential family with
\begin{equation}
 \label{eq:GLRM}
 \mathbb{E}(Y\mid \mathcal{X}(t), t\in T)=
 g\left(\alpha + \int_T\beta(t)  \mathcal{X}(t) dt \right),
\end{equation}
where $g(\cdot)$ is a monotone and smooth link function.

The model (\ref{eq:GLRM}) was extended by \citet{Goldsmith2015} to generalized multilevel function-on-scalar regression model which allows correlated errors considering random effects of subject specific variation and subject-visit specific variations.
For instance, assume that the functional response $\mathcal{Y}(t)$ follows a GFPCA (\ref{eq:GFPCA})
with subject and subject-visit effects, as in model (\ref{eq:twolevelmixed}),
and $p$ scalar predictors:
\begin{equation}
\label{eq:GFoSR}
\mathbb
\mathbb{E}[\mathcal{Y}_{ij}(t)\mid Z_i(t), U_{ij}] =
g\Bigl(
\beta_0(t) + \sum_{h=1}^p x_{hij}\beta_h(t) +
Z_i(t) + U_{ij}(t)
\Bigr),
\end{equation}
for subject $i$ at visit $j$.
\citet{Goldsmith2015} proposed to estimate this model expanding the random effects $ Z_i(t)$ and $U_{ij}(t)$ in their functional principal components, as in the two-level FPCA model (\ref{eq:twolevelFPCA}).

\section{Bibliographic search}
\label{sec:BibSearch}
The bibliography search in our study was done both on Web of Science (WoS, \citeauthor{WoS}, \citeyear{WoS}) and Scopus \citep{Scopus} along 2023 and 2024 (last access 26th of February, 2024).
Different combinations of keywords were used.
The search functions used in WoS and Scopus are listed in Table \ref{table:1}.

\begin{table}[ht!]
\begin{tabular}{l p{10cm}}
\hline
\textbf{Database}  & \textbf{Searching keywords}  
\\ \hline
 & \\[-.2cm]
WoS &
TS= (((``Functional data" OR ``Functional method" OR ``Functional PCA" OR ``Functional principal component") AND (``accelerometer" OR ``Wearable Device" OR ``biosensor" OR ``wearable sensor")) OR ((``Functional Data analysis" OR ``Functional PCA" or ``function-on-scalar regression") AND glucose))
\\
& \\[-.2cm]
& TS=(``Functional Data Analysis" AND ``heart rate" AND ``monitoring")
\\
& \\[-.2cm]
& TS=(biomechanical AND acceleration AND ``Functional data")
\\
& \\[-.2cm]
& \\
Scopus & TITLE-ABS-KEY (((((``Functional data")  OR  (``Functional PCA"))  AND  ((``accelerometer")  OR  ( ``Wearable Device")  OR  (``biosensor")  OR  (``wearable sensor")))  OR ((``Functional Data analysis")  OR  (``Functional PCA")  OR  (``function-on-scalar regression")  AND  glucose)))
\\
& \\[-.2cm]
& TITLE-ABS-KEY (``Functional Data Analysis" AND ``heart rate" AND ``monitoring") 
\\
& \\[-.2cm]
& TITLE-ABS-KEY(biomechanical AND acceleration AND ``Functional data") 
\\
& \\[-.2cm]
\hline
\end{tabular}
\caption{Keyword combinations used in bibliographic search.}\label{table:1}
\end{table}

As a result of this bibliographic search, a total of 155 works were found.  However, after manual revision 96 of these references were eliminated from the study due to many reasons such that the work was repeated, the study was not a full journal article (but was either a patent, grant, thesis, conference proceedings, abstract, or poster), the article was not related with FDA, the study was conducted on animals, or non-wearable devices were used in the study.
After this elimination finally 59 papers have been analyzed.

\begin{figure}
    \centering
    \includegraphics[scale=0.8]{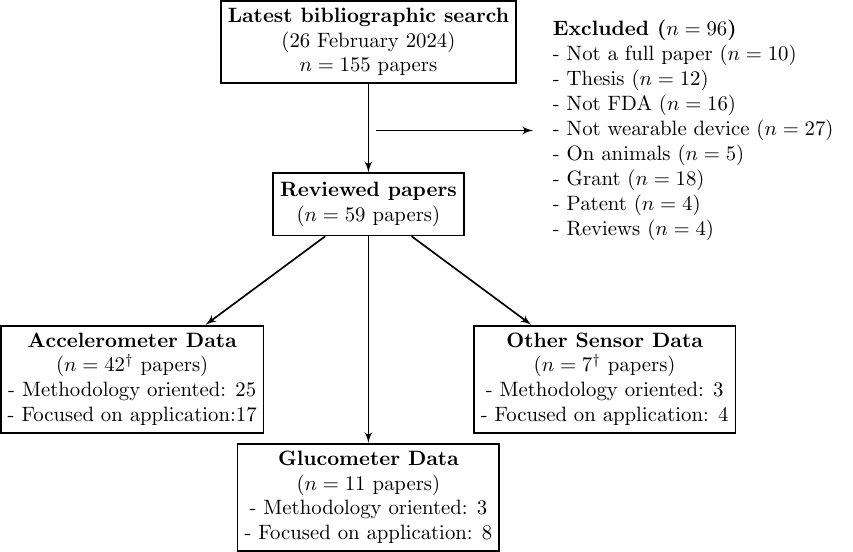}
    \caption{Bibliographic Search Process. $^\dagger$One article, \citet{Goldsmith2017}, includes both accelerometer and other type of sensor data.}
    \label{fig:bib_search}
\end{figure}

We summarize the bibliographic search process in Figure \ref{fig:bib_search}, taking into account the type of sensor data and whether the article contains methodological contributions or is focused on applications.

\section{Accelerometer Data}\label{sec:Accel}

Accelerometers are devices used to monitor any kind of movement, such as walking, intense physical activity, or even resting activity profiles.
Some accelerometers can also record heartbeats.
Accelerometer data is composed of the movement patterns of subjects monitored during a given time interval.
The continuous nature of the data allows it to be treated as functions of time.

In our review, we organize accelerometer studies into two subsections, according to whether they focus on a methodological contribution or an applied contribution.

\subsection{Methodological Contributions}
\label{sec:accmet}
The main two FDA approaches used in the articles related to accelerometers were FPCA and FLM (see Section \ref{sec:FDA}). Therefore, the methodological contributions will be handled in three subgroups considering the FDA approach in which the contribution was done (FPCA, FLM, or other methods).

\subsubsection{Contributions to FPCA.}
\label{sec:accFPCA}

Various novel FPCA methods have been proposed to deal with accelerometer data \citep{Goldsmith2015, shou2015structured,  xiao2015quantifying,
Johns2019, Wrobel2019, Backenroth2020, McDonnell2021, Zhong2022}.

\citet{Goldsmith2015} proposed generalized multilevel FPCA approach going beyond both the generalized FPCA \citep{HallMullerYao:2008} and the multilevel FPCA approach \citep{di2009multilevel}. This new framework allows a distinction between within-subject variability and between-subject variability when the functional data come from an exponential family distribution. Furthermore, the authors used Bayesian analysis to estimate the principal component scores and the coefficients of spline basis functions that are used to represent FPCAs. More detailed information on this work will be given in \ref{sec:accFLM} because this paper also concerns generalized multilevel functional regression.

\citet{shou2015structured} introduced structured FPCA as a novel dimension reduction method which is based on the spectral decomposition of the covariance operators that are estimated by symmetric sum method of moments as proposed by \citet{Koch1968}. This way, the authors aimed to simplify the dimension reduction process of crossed and nested designed multilevel models. The performance of the model was tested on the LIFEmeter accelerometer data used in the study of \citet{bai2012movelets}. The objective was to analyze the daily energy expenditure of the participants. The data was recorded at hours within days for each subject and therefore had a three-way nested structure (individuals, days and hours). For each individual there were at most 19 curves per day, each having length of one hour with measurements at minute level. The most important mode of variability ($76\%$) was found at hour level which indicates that energy expenditure of individuals were changing a lot during a day. The variation at the subject and day levels respectively accounted for $15.3\%$ and $8.3\%$ of the total variability.

\citet{xiao2015quantifying} analyzed activity data coming from  ``Baltimore Longitudinal Study on Aging'' (BLSA; see Section \ref{sec:Open} for a brief description of this study). The objective of the study was to analyze the circadian rhythms (that is, daily activity profiles) of subjects along age and across gender
in order to understand how subjects differ in activity and
to identify individuals who exhibited lower/higher activity levels compared to their respective age-specific groups.
The log activity counts of subjects are modeled as a function of both the time of the day and the age of individuals.
Two distinct models are fitted, one for each gender group.
After an initial bivariate smoothing, the residuals are analyzed using multilevel FPCA approach which accounts both for the variation between subjects (different individuals in the study) and the variation within subjects (different days of follow-up for the same individual).

\citet{Johns2019} introduced variable domain FPCA (vd-FPCA) to be used in the case that the functional observations are measured on an uneven domain. The proposed method was compared with two alternative methods: a ``scaled" method based on linear interpolation, and a ``weighted" method specifically proposed for censored data. The three methods were applied to a real accelerometer data set recording the movement of 20 subjects when they were standing up from a chair. It was concluded that vd-FPCA method was able to explain a high proportion of variation by using fewer components compared to the other FPCA methods.

\citet{Wrobel2019} proposed {\em registration} and FPCA for exponential family curves, which is an extension of Generalized Functional Principal Components (GFPCA)  \citep{HallMullerYao:2008}. The goal of registration (or alignment) was to set apart amplitude variability from phase variability in functional data. The proposed two-step estimation procedure was used to find FPCs for binary response variables. First, subject-specific mean functions $\mu_i(t)$ were estimated by GFPCA, and then {\em warping functions} (i.e., transformations of time units) were obtained by maximizing the log-likelihood of the exponential family distribution.
An application of this method has been done on binary activity curves obtained from BLSA data which show whether the participant was active or not at time $t$.
In this case, the smooth subject-specific mean $\mu_i(t)$ was the probability of subject $i$ to be active at each minute during 24 hours. Registration helped to analyze the phase variability and to determine the active/inactive periods of individuals.

\citet{Backenroth2020} proposed Non-negative and Regularized Function Decomposition (NARFD), which is an extension of non-negative matrix factorization (NMF; \citeauthor{LeeSeung1999NMF}, \citeyear{LeeSeung1999NMF}) to functional data. In NMF a non-negative data matrix is approximated by the product of two lower rank non-negative matrices. In NARFD non-negative functional data are expressed as sums of a few non-negative prototype functions. NARFD was applied to analyze Poisson distributed one-minute functional activity count data obtained from BLSA study. The decomposition provided by NARFD was compared with that coming from GFPCA (for which the authors proposed an alternative implementation for Poisson count functions). NARFD prototypes showed easy interpretation since they were positive in short time intervals. We consider that \citet{Backenroth2020} represent a contribution to the literature on exponential family FDA,
similar to the studies of \citet{HallMullerYao:2008},
\citet{Goldsmith2015} and \citet{Wrobel2019}.

\citet{McDonnell2021} combined curve registration and GFPCA (based on \citeauthor{Wrobel2019}, \citeyear{Wrobel2019}) to find out interpretable components that are able to differentiate subject-level variability in activity intensity from the subject-level variability in the timing of the physical activity. In the application, 24-hour accelerometric rest-activity profiles (including sleep and awake but not active periods) obtained from BLSA study were used. In the analysis, the minute-level activity-counts were transformed to ``activity probability profiles" which showed the probability of active state over time during a day for each subject. For the registration process, a two-step procedure was implemented which is a combination of warping functions and FPCA. In the first step, the warping functions were defined by using a two knot piece-wise linear function which leads to a series of three slopes. Registration of activity probability curves by using warping functions helped to reveal common activity patterns among subjects. In the second step, a binary FPCA approach was used to estimate the subjects' mean activity probability function over a 24-hour period. The warping function coefficients and FPCA coefficients were used to classify subjects according to their chronobiology,  which groups subjects according to the differences between their wake/sleep times and low/high activity probabilities.

\citet{Zhong2022} proposed Kendall FPCA,
an FPCA method for non-Gaussian data, and applied it to physical activity data.
The authors defined the Kendall's $\tau$ covariance function $K(s,t)$ as the covariance function of the functional random variable $(X(t)-\tilde{X}(t))/\|X-\tilde{X}\|$,
where $\tilde{X}$ is an independent copy of $X$.
It can be proved that $K(s,t)$ shares the eigenfunctions with the covariance function $C(s,t)$ of $X(t)$.
When $X(t)$ is far from Gaussianity, the standard estimation of $C(s,t)$ would be biased.
Thus, the authors proposed a U-statistic estimator of $K(s,t)$ that is asymptotically unbiased, leading to valid estimators of the shared eigenfunctions of $K$ and $C$ and, consequently, to a valid FPCA.
The data set used in the paper originally comes from the study of \citet{kozey2014changes} and consisted of 63 moderately overweight but healthy office workers.
An accelerometer (ActivPALTM, www.paltech.plus.com) measured the subjects' thigh angle and movement speed at five-minute intervals for three hours (one hour warm-up, followed by 30 minutes of vigorous exercise, 25 minutes of cool-down, then sedentary activities).
The raw data was used to estimate energy expenditure levels in metabolic equivalents (METs).
The energy expenditure levels were analyzed using the proposed Kendall FPCA.
The first two eigenfunctions had clear interpretations: the main variation occurs during the intense exercise period (due to the different physical training experience of the participants), and there is also a difference in the time at which individuals reach the peak of energy expenditure.

\subsubsection{Contributions to Functional Linear Models.}
\label{sec:accFLM}

Among FLM applications to accelerometer data, the most commonly used method has been Function on Scalar Regression (FoSR), where the response is a function and the explanatory variables are scalars \citep{Morris20061352, Goldsmith2015, Ghosal2021, Cui2022,javskova2023compositional}.
There were also methodological contributions related to Scalar on Function Regression (SoFR) with scalar predictors and functional response \citep{aguilera2020multi, tekwe2022estimation,jadhav2022function}
and
Functional Linear Concurrent Models (FLCM) where both the response and the predictors are functions \citep{Goldsmith2017, bai2018two}.

\citet{Morris20061352} aimed to analyze a case study that consists of weekday activity levels of children measured per minute by using Tri-Trac-R3D activity monitor, coming from the Planet Health study \citep{gortmaker1999reducing}.
A subset of 112 children was taken providing a total of 550 daily activity profiles recorded per minute from 9am to 8pm.
The objective of the study was to understand how the activity profiles are affected by various factors such as gender, age, weight, Body Mass Index (BMI), the average number of hours spent watching television per day, season (a factor showing whether the time is before or after the daylight savings time) and the school indicator (five schools were involved).
They used a Bayesian wavelet-based functional linear mixed model (BWFLMM) approach to model log-transformed activity profiles considering the children as random effects.
The activity profiles, as well as the functional coefficients, were represented by their coefficients in a wavelet basis expansion.
However, there were many incomplete profiles in the dataset.
Therefore, \citet{Morris20061352} proposed a data imputation method for functional data to complete activity profiles. First, they  fitted the BWFLMM using only the 95 complete profiles from 61 children. They then used the obtained posterior means of the model parameters to impute the missing data at each MCMC iteration when fitting the BWFLMM to the full data set.

In \citet{Goldsmith2015}, the main aim was to  estimate the effect of age and BMI on the probability of being active during  day time. They worked on a sample taken from BLSA study that consists of 583 subjects. The activity counts recorded per minute were converted to a binary variable which indicates whether the subject was active or not at recorded time. Then the probabilities of being active were computed over time.
A Bayesian generalized multilevel function on scalar regression model was proposed to analyze the relationship between scalar variables age and gender and the functional response of time varying probabilities.
The variations among subjects and among time were analyzed by a Bayesian generalized functional principal component analysis as mentioned in \ref{sec:accFPCA}.

\citet{Ghosal2021} developed a novel variable selection method for a nonlinear generalized FoSR model that allows the covariates to have both linear and nonlinear effects on the response. They applied the proposed method to accelerometer data taken from the 2003â€“2004 cohorts of the National Health and Nutrition Examination Survey (NHANES) \citep{leroux2018rnhanesdata}.
The goal of the study was to identify important associations between diurnal patterns of physical activity and demographic, lifestyle, and health characteristics of the participants. Generalized FoSR was used to model physical activity curves on categorical variables with linear effects and continuous predictors with nonlinear effects where the gender was handled as a confounding variable. The model revealed an important nonlinear relationship between physical activity and age of people. The levels of physical activity seemed to increase between ages 20-30 while was stabilized during midyears and started to decrease after age 60.

In \citet{Cui2022} a Fast Univariate Inferential (FUI) approach for FoSR was developed to make inferences for the fixed effects of a longitudinal functional dataset.
FUI consists of three steps: (1) fit a univariate pointwise mixed-effects model for each observed argument value of the functional response; (2) smooth the estimated univariate coefficients along the functional domain to obtain the estimated functional coefficients; and (3) use bootstrapping of study participants to make inferences (e.g., to obtain joint confidence bands for the functional coefficients).
The hip-worn accelerometer data of NHANES study, belonging to the period 2003-2006, were considered.
The objective of the study was to analyze the association between the functional response, which is a longitudinal indicator of sedentary activity (1 if activity count at point t exceeds 100 and 0 otherwise), and scalar predictors: gender, age, and day of the week.
In total 8 765 curves belonging to 1 680 distinct subjects were analyzed. They consisted of 1 440 observations per day. As a result, the novel FUI approach showed a good computational performance which was proved as well by a simulation study comparing the computational effort of FUI to two alternative models: functional additive mixed models (FAMM) and generalized multilevel FoSR \citep{Goldsmith2015}.

\citet{Goldsmith2017} discussed variable selection methods for the FLCM model with an application on Masked Hypertension Study \citep{schwartz2016clinic} where 24-hour systolic blood pressure levels were taken as a functional response and the individual measurements of physical activity, location, posture, mood, and other quantities considered as functional predictors. In the study, a variational Bayes algorithm was proposed to estimate the effect of coefficients.
Given that physical activity is not the primary variable in this work, we leave the detailed explanation to Section \ref{sec:Other} devoted to {\em other sensors}, since the response was measured by a 24-hour ambulatory blood pressure (ABP) monitoring device.

\citet{bai2018two} proposed a two stage model to analyze effect of both time-varying and time-invariant parameters on a count process conditional to a time dependent binary process.
The motivation behind the study was to identify whether a subject was active or not at a given time period (and his/her activity intensity when active) based on time-invariant physical characteristics (age, gender, BMI) and minute level Heart Rate (time-dependent variables) of the subject. In the study, a sample of 878 subjects from BLSA data were considered who have at least 3 full days of monitoring data. For each subject, just one day of data with the highest amount of physical activity count was chosen. For each subject, $Y(t)$ is the activity count observed at minute $t$, $A(t)$ is a binary process that takes value of 1 if the activity count of subject $i$ at minute $t$ is greater than zero, $Z_i(t)$ is the vector of time-independent covariates, and $H_i(t)$ represents the time-varying predictors.
In the first stage of modelling, for every minute $t$ the probability of $A_i(t)$ equal to $1$ was modeled on $Z_i(t)$ and $H_i(t)$ using a standard logistic regression model.
In the second stage of modelling, the logarithm of activity counts, $Y_i(t)$, was modeled on the same predictors $Z_i(t)$ and $H_i(t)$ for each subject conditioning on that the person is active at minute $t$ ($A_i(t)=1$), leading again to a different log-linear model for each minute.
In the estimation process, the time-variant coefficients were first estimated at these minute-specific models and then they were smoothed by LOWESS.
Regarding the time-invariant coefficients, the authors used two different approaches to combine their minute-level estimations.
As a result, both of the time-independent (physical factors) and time dependent (minute-level heart rate) covariates  were associated both with the activity count and the probability of being active. The activity intensity of subjects was significantly changing at different times of day for different gender and age groups.

In the context of multi-class classification,  \citet{aguilera2020multi} proposed a Functional Linear Discriminant Analysis (FLDA) approach based on penalized Functional Partial Least Squares (FPLS, \citeauthor{Preda_Saporta_2005_FPLS}, \citeyear{Preda_Saporta_2005_FPLS})  which is an optimization method based on the maximization of the covariance between the functional independent variable $\mathcal{X}(t)$ and the scalar response $\mathbf{Y}$.
The authors applied it to two different case studies, of which only the first one was produced by a wearable device. It consisted of biomechanical data recorded by smartphones coming from the study of \citet{anguita2013public}, where 29 subjects were monitored while performing three different types of movement: walking, climbing stairs and descending stairs.
The goal of the study was to determine the type of movement from the physical activity curves of the subjects. Since the movement type was a scalar response and activity functions were functional predictors, the linear discriminant analysis problem is considered as a SoFR model. Then, FPSL approach was used to fit the discriminant functions. The authors found that the functional approach performed better than the corresponding multivariate one: the correct classification rates in a test sample were 78\% and 56\%, respectively.

Recently, alternative ways to fit SoFR models were proposed to reveal associations between a scalar response and functional covariates. For instance, a two step estimation method that accounts for measurement error was proposed by \citet{jadhav2022function}.  The authors considered the case SoFR model where the functional covariate $\mathcal{X}(t)$ is not directly observable. The effect of this  unobserved variable was estimated by means of two surrogate functional variables: one instrument functional variable $\mathcal{Z}(t)$ and one contaminated version of the variable of interest with an additive measurement error $\mathcal{W}(t)$. In the first step the effect of functional instrumental variable on the unobserved functional variable was estimated by using a FLCM on $\mathcal{W}(t)$. In the second step, the effect of the unobserved variable $\mathcal{X}(t)$ on the scalar response was estimated using the coefficients obtained from the first step. The application of the proposed method was done on wearable device data obtained from NHANES data set (for the period 2003-2004) in order to analyze the relationship between physical activity curves and the BMI of subjects.

\citet{tekwe2022estimation} consider the same problem (SoFR with measurement error in a functional covariate).
The proposed model is constructed by regressing conditional quantiles of a scalar random variable $Y$ on a square integrable random function $\mathcal{X}(t)$ defined on the interval $[0,1]$ and on a vector of error-free covariates, $\mathbf{Z}$,  defined in $\mathbb{R}^p$. Such as in \citet{jadhav2022function}, it is assumed that $\mathcal{X}_i(t)$ is not directly observed.
The simulation extrapolation approach
\citep{cook1994simulation, carroll2006measurement}
for quantile regression with scalar covariates with measurement error was adapted to the case of functional covariates with measurement error and was used to estimate the model coefficients. This methodology was applied on accelerometer data of NHANES for the period 2005-2006 to understand the relationship between BMI and physical activity in adults aged between 20 and 85. The race, gender and age of the subjects were included as error-free vector valued covariates to the model while hourly averages of activity counts was handled as a functional covariate which is approximated by the functional instrumental variable of step counts. Four different percentile level of BMIs were used as response (25th, 50th, 75th and 95th quantiles). They compared results of the naive-based estimator approach to the error-corrected estimation approach proposed in the article. The findings have been supported by a simulation study. The results of the model adjusted for age, sex and race demostrated that the effect of physical activity counts on different quantiles of BMI varied depending on the time of the activity.

Recently, \citet{javskova2023compositional}
adapted ordinary compositional data analysis \citep{Pawlowsky-Glahn_et_al_2015} approach to SoFR. In the application, they used a dataset consisting the physical activity data of 74 girls whose age is between 14 and 17 \citep{gaba2021replacing}.
The physical activity intensities were collected at each five seconds by a tri-axial accelerometer worn to the wrist. The accelerometer data were handled as density functions defined in a Bayes space. Then by applying a centred log-ratio transformation, an isometric isomorphism was established between Bayes and Hilbert Space which allows to do operations between the elements of the Bayes space \citep{Egozcue_et_al_2006}. To smooth the probability density functions a novel method named ZB-splines, based on compositional splines, were introduced. The ordinary SoFR was extended to compositional scalar-on-function regression which is based on ZB-spline representation of the functional predictors which are defined as probability density functions. To reduce dimension and to find out the most important variations  they used Simplicial Functional Principal Component Analysis (SFPCA), an extension of FPCA for density functions, proposed by \citet{hron2016simplicial}.
 Hence, the functional parameter of the proposed regression model was obtained  by rewriting the basis expansions of transformed functions using SFPCA. Additionally, they proposed Compositional functional isotemporal substitution analysis (CFISA) to analyze the expected changes in the body fat percentage of the subject (adiposity) related to the specific subdomains of the time-based distribution obtained from accelerometer curves.

\subsubsection{Other Methodological Contributions.}
\label{sec:accother}

Other methodological contributions on functional data analysis of wearable device data were mainly related to clustering individuals:
\citet{Lim2019}, \citet{Jang2021} and \citet{song2023multi}.
However, there were also novel methods proposed
for fitting a Cox model on functional data \citep{Cui2021},
for inferential statistics \citep{chang2022empirical},
for functional kernel machine regression \citep{naiman2022multivariate},
and
for modeling changes in diurnal physical activity patterns based on
diffeomorphisms between Riemann manifolds \citet{Zou_et_al_2023_AnnApplStats_Riemann}.

\citet{Lim2019} proposed a two-step clustering method where the physical activity curves are first transformed either by using a standard rank-based transformation or the thick-pen transformation \citep{fryzlewicz2011thick},
a way to smooth raw data alternative to more conventional methods, such as spline or kernel smoothing.
The transformed functions are then clustered using the functional clustering approach proposed by \citet{Chiu_Li_2007_Funct_Clust}, a variant of the functional $k$-means method that considers both the mean and the modes of variation for each cluster.
In the application, the authors clustered individuals first according to their activity levels based on rank-based transformation of the input variable (low activity group, medium activity group, and high activity group) and then according to their activity patterns by using thick-pen transformation and obtained four different clusters. In the simulation study, the authors compared the performance of the proposed techniques to other functional clustering methods (FPCA, fclust, funclust, curvclust).

\citet{Jang2021} aimed to identify and group the daily activity patterns of 23 individuals using a clustering method with two novelties: {\em binning} and {\em ensembling}. In the study, step count data collected from the Fitbit wearable device (\url{https://www.fitbit.com}) between the years 2014 and 2015 were used. The step counts of 23 individuals were recorded each minute in a day between 12 AM to 11:59 PM giving a time series of 1 440 minutes a day per person.
The authors first proposed to bin each individual's daily count data by transforming it into a histogram, with a chosen bin width and starting point.
Then the set of histograms was clustered by $k$-means for the selected $k$ according to the gap statistic.
This process was repeated for different combinations of bin width and starting point, leading each to a different clustering result.
In the final ensemble step, these results were aggregated by a voting scheme proposed by \citet{dimitriadou2002combination}.
In a simulation study, the performance of the new method was compared with the performance of three existing clustering approaches: curvclust \citep{giacofci2013wavelet}, funFEM \citep{bouveyron2015discriminative} and $k$-means (considering as well the ensemble versions of funFEM and $k$-means approaches).

\citet{song2023multi} proposed a novel clustering algorithm for multivariate functional data which comes from the step count data obtained from the pedometer Fitbit armband. The data consists of step count data at minute level for 79 subjects. In total 21 394 days were observed.
The objective of the application was to cluster days.
Classical clustering methods are not well suited for step data because they are discrete and zero-inflated.
Thus, the authors proposed a new clustering method that was based on generation of new functional variables (namely, the cumulative sum function, the ordered slope function, and the mean score function) based on the features of quantity, strength and pattern of the observed functional data and then to use multivariate functional principal component analysis (MFPCA).
They applied classical $k$-means \citep{macqueen1967some} and partitioning around medoid (PAM, \citeauthor{kaufman2009finding}, \citeyear{kaufman2009finding}) algorithms on the MFPCA scores.
In both cases the optimal number of clusters (according to the gap statistic) was equal to 7, the observed days were quite evenly distributed among clusters (specially for PAM) and the clusters had clear interpretations (some of them being related with weekends).

In the study of \citet{Cui2021}, a non-parametric additive functional Cox model approach was introduced to analyze the relationship between physical activity functions and time-to-death of subjects from NHANES study from the period 2003-2006.
This model extends, on the one hand, the functional generalized additive model \citep{mclean2014functional} to allow modeling of time-to-event data and, on the other hand, the functional linear Cox regression model, developed independently by different authors
\citep{gellar2015cox,
lee2015bflcrm,
Qu_et_al_2016FunctCoxModel,
kong2018flcrm},
to relax the functional linearity assumption.
The functional predictor in the model was the log-transformed daily activity curve of individuals (averaged over the observed days and smoothed by FPCA) and the response was survival time.
The functional model was adjusted by factors based on the subjects' socio-demographic status, health status, and disease indicators.
The results indicate that, regardless of age, there is an increased risk of mortality for individuals who are above the 60th percentile of nighttime activity and below the 35th percentile of daytime activity for their age group.

\citet{chang2022empirical} proposed a non-parametric empirical likelihood-based inference method, which does not require smoothing, for comparing functional means and constructing confidence bands for functional data that are bounded, non-increasing, and possibly discontinuous.
Their approach uses the framework of stochastic processes having
right-continuous sample paths of bounded variation.
They applied this approach to NHANES data from the period 2005-2006 to compare mean {\em occupation time curves} obtained from raw activity data corresponding to subjects aged 65 and older when they are divided into four groups according to the veteran status (yes or no) and age (older than 75 or not).
For a certain activity profile $X(t),\, t\in [0,1]$, the corresponding occupation time curve $L$ evaluated at a given activity level $a$ (expressed in counts per minute) is defined as the proportion of time the $X$ spends above $a$. Then the function $L$ is bounded, non-increasing, and right-continuous, as required in the theoretical results developed by the authors.
The occupation time curves $L(a)$ were defined for $a$ in the interval $[0,499]$, which is considered in the literature to be the range of sedentary activities.
The effect of age was found to be significant, regardless of veteran status, while the effect of veteran status is significant only for the older age group.

\citet{naiman2022multivariate} introduced the functional kernel machine regression (FKMR) model which combines the functional framework with machine learning regression (specifically, the least squares kernel machine method, LSKM, \citeauthor{Liu2007_LSKM} \citeyear{Liu2007_LSKM})
to explain a scalar response from multiple functional predictors and possibly other scalar covariates.
The motivating real problem was to understand the relationship between obesity and the physical activity in children. The authors used a data set coming from the Early Life Exposure in Mexico to ENvironmental Toxicants (ELEMENT) study \citep{lewis2013predictors} that includes tri-axial accelerometer data of 395 children recorded by a wrist worn device and physical characteristics of children.
A scalar-on-function regression model were fitted between the minute level activity count functions of the children (at three different axes) and BMI which is an indicator of obesity in children.
The FKMR fitting process proposed in the paper has two main steps.
First, the functional predictors are represented by the FPC scores at each individual in the sample.
Then, a LSKM is fitted using these scores as predicting variables, reducing the SoFR model to a non-functional additive regression model.
A sparse group lasso regularization term is included in the LSKM objective function in order to perform simultaneously model fitting and variables selection (which functional predictors are relevant, and which of their principal functions are useful).
A simulation study showed that the proposed FKMR algorithm performed better than other alternative methods.
Similar results were obtained in the BMI real data problem,
where the FKMR had an adjusted $R^2$ of $0.30$, nearly twice as big as other competing models.

The main objective of \citet{Zou_et_al_2023_AnnApplStats_Riemann} was to model changes in physical activity diurnal patterns through a clinical trial, in order to quantify how they are affected by life-style interventions, and how they are related with health outcomes.
This study was motivated by two clinical trials:
the MENU 12-month study \citep[Metabolism, Exercise and Nutrition]{MENU}
and
the RfH 6-month study \citep[Reach for health trial on cancer patients]{RfH}.
Both studies involved overweight women and had weight loss as the main health outcome.
The authors proposed to characterize an smoothed physical activity record as a one-dimensional Riemann manifold (a smooth curve) embedded in $\mathbb{R}^2$ (the first coordinate for time of the day, and the second one for the Euclidean norm of the triaxial physical activity records).
Given two different physical activity profiles for the same subject (the first one corresponding to the beginning of the clicinal trial, and the second one to the end), the change from one to the other is modeled as a subject-specific diffeomorphism (a bijective differentiable function, with differentiable inverse) between Riemann curves.
Assuming that this diffeomorphism is the result of small changes that smoothly affect the physical activity patterns from the beginning to the end of the clinical trial, and using concepts from ordinary differential equations and reproducing kernel Hilbert spaces, the authors proved that there is a unique subject-specific minimum deformation energy diffeomorphism that transforms the initial activity pattern into the final one, and that it is determined by the initial deformations of the minute-level points in the initial curve, which are vectors in $\mathbb{R}^2$ called {\em initial momenta} and stored as a subject-specific matrix of dimension $m\times 2$, where $m$ is the length in minutes of the studied period of the day.
A standard principal component analysis  of initial momenta allowed the authors to discover the main modes of variations between subjects in changes of physical activity patterns. Moreover, the individual principal component scores were used as responses or as predictors in different multiple linear regression models.
In the MENU study a total of 177 subjects (overweight non-diabetic women) had both baseline and month-12 physical activity records available. It was found that 4 principal components of the initial momenta were significantly associated with change in BMI, suggesting that an increase in physical activity at any time of the day (and especially in the morning) from baseline to month-12 is associated with a decrease in BMI).
From the RfH study (with records of 303 overweight, postmenopausal early-stage breast cancer survivors women), the authors concluded that the life-style intervention (following a weight loss program) may have had a positive effect in promoting daily activity, with a shift in physical activity to later hours.
The paper also included a simulation study and a comparison with the time warping method (or registration, based on \citeauthor{Wrobel2019}, \citeyear{Wrobel2019}).

\subsection{Applications of FDA on Accelerometer Data}
\label{sec:accapp}

The applied contributions on accelerometer data are mainly based on FPCA and FLRM approaches.

The main objectives for using FPCA in the relevant studies \citep{Zeitzer2018,Xu2019, yeung2023us, winer2024impaired} was to reduce the dimension and to find out the most important modes of variations of activity profiles. Also in most of the studies an ordinary multivariate regression model was constructed between obtained FPC scores and the related response to understand the associations between variables of interest.

In \citet{Zeitzer2018} FPCA was applied to the daily activity patterns of 2 976 males participating in the Osteoporotic Fractures in Men Study (MrOS).
They were recruited between the years 2000 and 2002  in various clinical centers, and followed up from 6.5 to 8 years.
In the study the subjects were monitored using a wrist-worn accelerometer (Octagonal SleepWatch-O, \url{https://www.ambulatory-monitoring.com/}) placed on the non-dominant wrist.
The variation patterns of physical activity functions were explained by means of FPCA.
The first four components explain 88\% of the variability and each of them is associated with a specific activity pattern.
The authors first explained the FPC scores using multiple linear regression with various psychological and demographic characteristics of subjects as explanatory variables.
Then the FPC scores are used as predictors in different models to explain changes in sleep quality, cognitive function and mortality, adjusting for demographic and socio-economic variables.
The authors concluded that daily activity patterns may be useful as predictive biomarkers for these outcomes.

In the study of \citet{Xu2019} the effects of obesity and energy expenditure on cancer risk for women were analyzed.
The sample was constructed by unifying 245 overweight healthy women in the MENU study \citep{MENU}
and 333 overweight breast cancer survivor women from the RfH study \citep{RfH}.
The multilevel FPCA approach based on mixed models \citep{di2009multilevel} was used to analyze temporal variation in the physical activity curves of subjects.
The authors proposed to use Euclidean norm of the individual-level principal component scores as a measure of the total variation of daily activity data to model the relationship between cancer status and activity patterns.
Thus, they would be able to interpret the differences and peaks in activity functions of women with the same average activity levels.
To understand how physical activity patterns change with health status, several t-tests were performed to compare cancer and control groups on FPC scores for the first four components.
The authors concluded that the cancer group tended to have lower morning activity.
Moreover, the relationships between physical activity patterns (summarized by FPC scores) and different health outcomes based on quality of life and different biomarkers were analyzed.
Specifically, for the binary outcome insulin status (sensitive versus resistant) a logistic regression model was fitted,
while for quantitative biomarkers linear regression models were used, both adjusted by BMI and cancer status.
In this way, they were able to analyze the effect of physical activity on parameters related to health at different times of the day, leading them to confirm the health benefits of physical activity.

\citet{yeung2023us} investigated the association between the rest activity profile of adolescents (aged between 12 and 19) and their demographic and socioeconomic status.
The 24-hour activity records of 1 814 adolescents who participated in NHANES between the years 2011 and 2014 were treated as functional data. The mean of the Monitor Independent Movement Summary \citep{john2019open}
at five-minute epochs was used to measure activity.
Applying FPCA to the rest activity profiles, the most important variations of the daily activity data were analyzed by the first four components which account for 82.7\% of the total variation and are associated respectively with high amplitudes of activities, early time activities, early activity peaks, and prolonged activity.
Then, multiple linear regression models were applied to explain FPC scores in terms of personal characteristics (age, gender, ethnicity, household income, and parental education).
The differences between weekends and weekdays were also analyzed.
These models suggested broad influences of demographic characteristics and family socioeconomic status on physical activity in adolescents.

The main objective of the \citet{winer2024impaired} was to analyze the relationship between activity level and the development of Alzheimer's disease (AD) and Parkinson's disease (PD) and cognitive decline.
Diurnal accelerometer data of 82 829 subjects from UK Biobank study between June 2013 and January 2016 were considered in the study.
Participants were followed for several years (mean follow-up 6.8 $\pm$ 0.9 years), with 187 individuals diagnosed with AD and 265 with PD.
Two sets were defined with matched controls who did not progress to AD or PD (age, sex, education, general health, BMI, and deprivation index were used to determine matches).
Two different FPCAs were performed (one for diagnosed with AD and their matches, and another analogous for PD) where the functional data were
smoothed individualâ€™s 24-hours median accelerometer data.
Mann-Whitney test was applied to the scores of the first four components
of subjects with diagnosis and their matched controls.
The authors found a significant difference between individuals who progressed to PD and who did not progress PD in the first component scores, which was related to the amplitude component.
This led to the conclusion that PD predisposition is associated with lower daytime activity.
On the other hand, there were no significant differences between those who progressed to AD and matched controls.

The applied contributions in FLRM can be considered in three groups depending on the type of the response variable and the type of the predictors in the model. The first group of FLRM was based on FoSR and FANOVA models \citep{Goldsmith2016, Nieto-Reyes2017, Sera2017, Wu2019598, Rackoll2021, Wrobel2021, lin2022longitudinal, matabuena2022estimating}.

\citet{Goldsmith2016} used functional linear models on 24-hours diurnal profiles of 420 children obtained from accelerometers. FoSR models were used to reveal the association between timing of physical activity and various covariates (season, sex, BMI z-score, presence of an asthma diagnosis, behaviourial variables of child, and mother's demographics).
The results obtained from FLRM were compared to the results of formerly applied multiple linear regression (MLR) models. The effect of mother's birth place was found significant on activity level of the child in FoSR models while it was not found important in MLR. This way, FDA approach contributed to the interpretation of the model.

\citet{Nieto-Reyes2017} analyzed accelerometer data obtained from the sensor of Android smartphones with the aim of classifying Alzheimer's patients according to the stage of the disease (early, middle and late stages).
Different FANOVA methods \citep{cuesta2010simple,zhang2019new} were applied for each axis ($x$, $y$, and $z$ axes) to compare accelerometer curves of 35 patients who are in different stages.
In general, the null hypothesis of equality of the mean accelerations for the different stages of the disease was rejected.
Significant differences were found between early and intermediate stage patients and between intermediate and late stage patients.
Moreover, a functional supervised classifier based on functional depth (for more information on functional depth see, for instance, \citeauthor{Febrero2008}, \citeyear{Febrero2008})
and non-parametric kernel approach was proposed to classify patients,
following \citet{cuesta2017dd}.
The classification results were not satisfying, with correct classification rate around 50\%.

\citet{Sera2017} aimed to analyzed the physical activity profiles of school-aged children in UK, in relation with time effect, geographical, demographic, and behavioural characteristics.
They analyzed 6 497 children from the Millennium Cohort Study \citep{smith2002millennium}, a prospective study of children's social, economic, and health conditions in the UK between September 2000 and January 2002.
For each child, only one randomly chosen daily profile was considered.
A multivariable FANOVA model was used to analyze the effect of a total of 13 distinct factors. Bootstrap was used to provide confidence intervals to the estimated functional effects.
The authors conclude that physical activity patterns change significantly
according to gender, ethnicity, region, season of measurement, social activity, mode of travel to the school, and number of cars in the household.

To analyze the association between physical activity and sleep quality of post-partum women, \citet{Wu2019598} used a descriptive FDA approach on 24-hour physical activity curves recorded over a week.
The sleep quality of the participants was measured by the Pittsburgh Sleep Quality Index (PSQI) whereas the 24-hour physical activity data were measured by a tri-axial accelerometer worn on non-dominant wrist.
The considered dataset consisted of 294 women with completed PSQI and six-month accelerometer data.
To measure the intensity of the physical activity Euclidean Norm Minus One (ENMO) metric at five-second epochs were used.
ENMO metrics were converted to continuous functions by using B-splines smoothing technique.
The PSQI was transformed into a dichotomous variable with categories for good and poor sleep quality.
Differences in physical activity during the day between the two sleep quality groups were tested using a FANOVA based on permutation tests, which only reached statistical significance briefly around noon.

\citet{Rackoll2021} used FoSR to analyze activity patterns of patients with mild cognitive impairment (MCI) and age-matched healthy older volunteers (HOV) coming from two different clinical trials: ``Effects of Brain Stimulation During Nocturnal Sleep on Memory Consolidation in Patients With Mild Cognitive Impairment" and ``Effects of Brain
Stimulation During a Daytime Nap on Memory Consolidation in Patients With Mild Cognitive Impairment". They used FoSR model by taking activity counts as the functional response and the group indicator as the explanatory variable. They compared the results of the functional model with the multiple linear regression model adjusted by age and gender where the response was the global average activity count during the day.
They found that MCI patients had periods during the day with lower activity levels compared to HOV. These differences were not noticed when the global amount of activity was used.

\citet{Wrobel2021} applied curve registration and functional regression models on 24-hour activity profiles for 88 793 adults
(39 255 men and 49 538 women, aged between 42 and 78)
from the UK Biobank accelerometer study to understand how physical activity patterns vary across ages and by gender.
The focus was on weekday activity patterns measured in ENMO units, which are condensed into a single physical activity profile in two ways: a continuous one (the average daily profile, with a resolution of one minute) and a binary one (for each minute of the day, an indicator of whether the majority of days have activity in that minute above a threshold used previously in the literature).
The binary activity profiles were aligned, and the warping functions were also modeled since they provide information about timing of physical activity.
Generalized FoSR were used to model each of the three functional responses (continuous ENMO functions, functional binary activity indicator, and warping functions) as smooth functions of time of the day, age, and gender.
The authors conclude that both the daily pattern of physical activity and the likelihood of being active change significantly with age. There are also significant gender differences in these trends.

\citet{lin2022longitudinal}
analyzed the association between diurnal physical activity and longitudinal health outcomes of 245 non-diabetes overweight women who participated in the MENU study \citep{MENU}.
First, the authors decomposed the daily activity profiles by using
a longitudional FPCA \citep{greven2010longitudinal}
taking into account both the subject-specific variability and the nested structure of data.
Then, the authors fit two regression models aimed to explain health indicators from the results of the longitudinal FPCA.
In the first model, the health variables were regressed on the functional principal components scores.
The second model is a FoSR where the functional effects of individuals and visits (both coming from the previous longitudinal FPCA) are functional predictors.
Both models were adjusted by age, ethnicity, smoking status, and follow-up visit of subjects.
Key findings include the following. More physical activity is associated with better metabolic health, which is accentuated when physical activity increases between visits (within an individual).
Moreover, physical activity earlier (rather than later) in the day is more beneficial for reducing weight.

Finally, \citet{matabuena2022estimating} intended to introduce athletic training researchers to analyzing repeated-measures functional data with multiple levels of aggregation.
The authors provided an accessible introduction to multilevel functional data analysis, and they made available a GitHub repository with the data and the R code used to perform all analyses.
The study focused on multilevel FPCA and multilevel functional regression models which are used to analyze biomechanical functional profiles of 19 healthy runners obtained during medium intensity continuous run (MICR) and high intensity interval training (HIIT) sessions.
Each functional data consists on the knee location trajectory (in one of three dimensions, $x$, $y$, and $z$) collected from one running stride as a function of time, which was standardized to the interval $[0,1]$.
There were 20 strides for each training intensity degree and runner.
The authors found a significant session type effect (MICR versus HITT) but the gender effect was not significant.

Other studies used Scalar on Function Regression (SoFR) to estimate a scalar response, treating the activity profiles as functional predictors \citep{Augustin2012, Augustin2017, franchi2023wearable}.

\citet{Augustin2012} proposed to use functional models to predict fat mass as a function of weekly physical activity profiles measured by Actigraph accelerometers. The data set composed of 10 080 observations per subject recorded through a week. The histogram of accelerometer counts were used as a predictor by estimating the contribution of each part of the histogram on fat mass as a smooth function. Generalized regression of scalars on functions was used to model response fat mass at age 12 and 14 years by using different predictors such as confounders (sex, height in meters, and squared height),
accelerometer weartime and the smooth function of physical activities that represents the contribution of accelerometer counts. In a novel approach proposed by \citet{Augustin2017} the scalar response  ``fat mass" was regressed on additive multidimensional functional predictors. The functional response was total fat mass curve of each individual, the predictor was the accelerometer profile for each individual. Three models were fitted considering different age groups of individuals. The first two models were based on confounders, weartime and physical activity at age 12 and 14 respectively. The third model was based on confounders and wear time at age 14 but physical activity at age 12.

Another applied study on accelerometer data was \citet{franchi2023wearable}. In this study, the motor behaviour of infants with brain damage have been followed by using wearable sensors. The objective of the study was to investigate the relationship between the movement data obtained from accelerometers and the scores obtained from two clinical scales: the Infant Motor Profile (IMP) score and Alberta Infant Motor Scale (AIMS). The sample consisted of 17 infants with a brain injury who were hospitalized in the Neonatal Intensive Care Units of three different University Hospitals of Italy (Santa Chiara Hospital, Pisa, Meyer Children's Hospital, Florence and Careggi General Hospital, Florence).
While playing in a controlled environment, the infants wore a tri-axis accelerometer on their wrists and trunks by custom bands. Four different functional predictors were defined by the authors based on the activity index \citep{bai2016activity}:
three hands' asymmetry indexes and one measure of the total movement.
The functional curves were aligned considering the start of the training for each subject. Four different scalar variables were considered as responses: AIMS total, IMP total without adaptation domain, IMP symmetry, IMP fluency and IMP performance.
Several functional linear regression models were constructed to analyze the relationship between clinical assessment scores and the functional predictors.
The results showed significant relationships between clinical outcomes and physical activity recorded by the sensors, indicating that functional linear models could predict clinical scores.

There is only one applied study focused on FoFR models, where both the predictor and the response are functions.
In \citet{Benadjaoud2019} the relationship between lung functions obtained during five seconds from a portable spirometry (ndd Easy on-PC Spirometer, \url{www.nddmed.com}) and the physical activity functions obtained from a triaxial accelerometer GENEActiv (\url{https://activinsights.com}) was investigated in terms of a FoFR model.
Activity data were collected from 3 063 subjects for 24 hours on seven consecutive days. The recorded acceleration was averaged in five-second epochs, and then the density function of the log-transformed acceleration records was estimated using a kernel estimator. The functional regression model adjusted for age, sex, ethnicity, height and weight was used to estimate the expired air volume-time function (during five seconds) of the subjects on their physical activity density. The results showed an association between physical activity and lung function in smoking and non-smoking adult groups. Specifically, for non-smokers, the association was more evident using the FDA approach compared to models using only aggregated figures.

Recently, the application of methods that combine machine learning with functional data analysis are appearing.
For instance, \citet{White2022} applied machine learning models based on FPCA to predict the peak power of jumping performance of 69 male and female athletes (who performed 696 vertical jumps in total) from bodyâ€™s vertical acceleration and velocity recorded by three inertial sensors attached to the different parts of the skin of the athletes (lower back, upper back and lower shank).
The information received by the sensors was converted to
accelerometer data.
FPCA was applied to extract important features from acceleration curves.
The performance of three different machine learning models (regularised linear regression, support vector machine, and Gaussian process regression) were compared by taking FPC scores as predictors and the peak power of jump of the athletes as response.
The authors conclude that the best predictions are obtained when using a finely tuned support vector machine, working with data coming from the sensor placed on the lower back.
Nevertheless, the level of accuracy of this model (an error around a 5\% of the mean peak power) does not meet the level of accuracy needed for practical use (the target error level was fixed at $3.4$\% in the literature).

\section{Glucometer Data}\label{sec:Gluco}

{\em Glucometers}  are devices designed to continuously monitor blood glucose levels (BGL) by measuring interstitial glucose levels through a sensor placed under the skin.
We found a total of eleven articles using glucose monitoring data with FDA approaches.
Four of these articles \citep{Law2015,Law2019, Scott2020, sibiak2024functional} had several authors in common and in all of them  FANOVA is applied to understand the effect of qualitative predictors on smoothed glucose profiles.
In the studies of \citet{gecili2021functional} and \citet{Gaynanova2022} FPCA was the main approach. Other three articles \citep{matabuena2021glucodensities,diazrizzolo2022,sergazinov2022case} were using functional linear models with the aim of making predictions where the glucose curves were handled either as a functional response or as a functional predictor.
\citet{Cui_et_al_2023_glucodensity} performed clustering analysis on estimated density functions of glucose concentrations.
Finally, \citet{matabuena2023reproducibility} had a different objective than the previous studies because it focused on the reproducibility of the continuous glucose monitoring (CGM) data.
All the articles, except four of them, were using Medtronic CGM systems (Medtronic MiniMed, Northridge, CA, \url{www.medtronic.com}) wearable device, \citet{Gaynanova2022}, \citet{sergazinov2022case} and \citet{Cui_et_al_2023_glucodensity} were using Dexcom G4 and G6 devices (\url{www.dexcom.com}) while \citet{diazrizzolo2022} were using Free Style Libre  (\url{www.freestylelibre.es}).

In \citet{Law2015}, \citet{Law2019}, \citet{Scott2020} and \citet{sibiak2024functional} the main objective was to understand the relationship between temporal glucose variation and the development of large-for-gestational-age (LGA) infants in women with treated gestational diabetes mellitus (GDM).
The glucose concentrations were monitored every five minutes, generating 288 measurements per day.
In \citet{Law2015} and \citet{Law2019} Continuous Glucose Monitoring (CGM) has been analyzed separately for each trimester during the pregnancy.
Prior to using the FDA approach, glucose levels evolution on time were summarized by 18 different measures, and then a total of 18 multiple regression analyses were conducted using LGA indicator as factor, adjusted by study center and type of diabetes.
The use of FDA allows to fit just a unique functional regression model, where representation of glucose levels as functions was done by a B-spline approach. This way temporal differences among diurnal glucose profiles were found across LGA and non-LGA groups, adjusted by study center and type of diabetes, for each trimester of pregnancy.

The aim of \citet{Scott2020} was to compare temporal glucose profiles of women with different characteristics. The study consisted of three comparisons.
First, real-time CGM was compared with standard self-monitoring and significantly lower glucose levels were found in CGM.
Second, the BGL of women using insulin pumps was compared with the BGL of women using multiple daily insulin injections. It was found that women using pumps had significantly higher glucose at six months of pregnancy, but no significant differences were found at the end of pregnancy.
Finally, the authors compared the BGL of women who had LGA infants with those who had non-LGA infants such as in the previous studies and concluded that the first group had significantly higher glucose levels.

In \citet{sibiak2024functional} the main objective was to compare the daily glycemic profiles of mothers of LGA and non-LGA newborns.
The subjects were selected among 545 patients with Type I diabetes (T1D) who had a visit in the Department of Reproduction of Poznan University of Medical Sciences in Poland between the years 2017 and 2021.
The sample consisted of 102 pregnant patients between the ages of 18 and 45 with a history of T1D for at least 12 months.
A sensor-augmented insulin pump was provided to the patients. The sensor recorded data every five minutes giving in total 288 measurements per day. By applying a FDA approach, similar to that used by
\citet{Law2015}, \citet{Law2019} and \citet{Scott2020},
the most significant differences in glucose profiles were found in the third trimester between the 26th and 32nd weeks of pregnancy.

\citet{gecili2021functional} analyze the BGL curves of 443 Type I diabetes patients \citep{juvenile2008continuous} using sparse FPCA \citep{yao_et_al:2005:sparseFDA} with two main objectives. First, they look for different {\em phenotypes}, defined as clusters of patients with similar BGL curves. To do so, they
divide individuals by the first and third quartiles of their scores at the first and second functional principal components.
This way they define 9 phenotypes of Type I diabetes with clear interpretation inherited from that of the principal functions.
Second, the authors want to analyze the joint temporal variation of BLG as a function of both, the hour of the day and the day of the week. A two stage FPCA, known as ``double FPCA approach" \citep{chen2012modeling}, was applied to analyze daily glucose value changes and their evolution at the first week of the study.

\citet{matabuena2021glucodensities} proposed nonparametric regression approaches for modelling density function of glucose levels (glucodensities) obtained from the sensor Enlite.
After converting glucose levels to glucosedensities by using a  Nadaraya-Watson density estimator \citep{Silverman_1986}
two different types of modelling were applied on glucodensities. The first model aims to predict from glucodensities various biochemical measurements such as mean amplitude of glycemic excursions (MAGE; glycemic excursions are fluctuations in BGL caused by meals) or mean of daily differences (MODD) in glucose concentration.
In this part of the study, non-parametric kernel functional regression models \citep{FerratyVieu2006} with the 2-Wasserstein distance between densities were proposed considering glucodensities as predictors.
The objective of the second model was predicting glucodensity functions on five biomarkers, including MAGE and MODD. To predict this functional response an adaptation of Fr\'echet regression \citep{petersen2016functional} was proposed by using 2-Wasserstein distance in the model.
As a result, it was found out that the glucodensities were good at predicting biomarkers. However, when the biomarkers were used as predictors the glucodensities could not be estimated with high accuracy.

\citet{Gaynanova2022} investigated the BGL fluctuations during sleep. The glucose levels of 124 subjects were monitored by Dexcom G4 sensor (www.dexcom.com) taking records every five minutes for seven consecutive days. Philips Actiwatch was used to determine sleep periods of individuals. For sleep periods of each individual glucose trajectories were obtained. The point-wise mean and standard deviation of BGL for each subject $i$ at time $t$ were smoothed by using FPC basis functions, letting $\hat{\mu}_i(t)$ and $\hat{\sigma}_i(t)$ be the FPC-based smoothed functions.
Considering two types of variability that exists between subjects and between the sleep periods of the same subject,
BGL is modeled as a multilevel functional model which assumes that
BGL at time $t$ for individual $i$ and sleep period $j$ follows
location and scale transformed Beta distribution with four parameters: the subject specific mean $\hat{\mu}_i(t)$, subject specific standard deviation $\hat{\sigma}_i(t)$, daily minimum BGL value $m_i$ and daily maximum BGL value $M_i$ where $[m_i,M_i]$ is the support of the distribution.
This method allows to compute fitted median function, 2.5\% and 97.5\% point-wise quantile functions of subject-wise BGL, which show an
asymmetric distribution for certain individuals.
Additionally, the scores obtained from FPCA for the subject specific mean and standard deviation were used with two aims.
First, the relationship between FPC scores of mean and those of standard deviation were analyzed and positive correlations were found, which indicates that participants who have higher BGL also have higher variability. In particular, both the mean and standard deviation of glucose value  decrease over the night.
Second, both sets of FPCA scores were used to predict glycosylated hemoglobin (HbA\_1c) levels of subjects.

In the study of \citet{diazrizzolo2022} the objective was to analyze the relationship between glucose levels and nutrient intake. In particular, the effect of quinoa on BGL was the main goal of the paper. The glycemic measurements recorded by Free Style Libre sensor were handled as a function of time. Due to the irregularity of the glucose curves, only the time interval that begins half an hour before the start of breakfast and ends two hours later was considered.
In the smoothing process warping functions were used to accommodate a possible measurement error in the reported breakfast starting time.
FoSR models were used to analyze the effect of consumed nutrients on the glucose functions during eight weeks period where in the first four weeks the subjects follow a regular diet (RD) and then in the latter four weeks a quinoa-based diet (QD). The explanatory variables included to the FoSR model were indicator of diet type, patient indicator
and the contents of different nutrients. Various models were applied from simplest to most complicated. In the simplest model, only the diet factor was considered as a predictor. In the second model both the diet and the patient factors were taken as predictors. Then the scalar nutritional parameters were included into the model. In total 132 nutritional variables were considered where 89 of them were found significant.
After removing the non-significant nutrients a more complex model has been constructed with the diet factor, patient factor and
40 nutritional variables treated as constant coefficients, and other 49 treated as functional coefficients. The functional regression models showed that the consumption of a diet rich in quinoa has a reducing effect on blood glucose fluctuations compared to the usual diet.

The case study of \cite{sergazinov2022case} considered the CGM values during sleep of 174 patients with Type II diabetes mellitus obtained from the HYPNOS study \citep{rooney2021rationale}.
The main objective of the article was to analyze the effects of diabetes medications and obstructive sleep apnea (OSA) on glucose levels.
Then, FAMM and FUI methods were used to estimate the multilevel function-on-scalar regression model and to construct pointwise and joint confidence intervals for the inference (the last are available only when using FUI).
The results obtained from both methods were compared to understand the effects of several covariates (age, sex, BMI, HbA\_1c, OSA severity, Biguanide and Sulfonylurea) on the glucose levels.
The main conclusion was that moderate-to-severe OSA is significantly associated with higher glucose trajectories during sleep.
From a methodological point of view, they conclude that FUI is more flexible and faster than FAMM.

\citet{Cui_et_al_2023_glucodensity} focused on an unsupervised classification problem by clustering the density functions of CGM values recorded at each five minutes. The considered data was coming from an ongoing clinical trial \citep{SoSaE} on pediatric and young adult subjects with Type I diabetes aged between 12 and 21 years old.
The number of daily curves recorded for the 30 subjects included in the analysis ranged from 3 to 468, with an average of 84.7 curves per subject.
First, the glucodensity functions were estimated as previously proposed in the study of \citet{matabuena2021glucodensities},
resulting one estimated density function per patient.
The authors used the R package {\tt{biosensors.usc}} 
\citep{biosensors:2022}
to do clustering based on Wasserstein distance between between glucodensities and a generalization of $k$-means called $k$-groups  energy distance clustering \citep{energy_R:2022}.
In total three clusters were obtained. The first group consisted of six participants with the highest average and highest variability of glucose levels, the second group consisted of 11 individuals with a better glycemic control and finally the last group consisted of 13 subjects with the lowest average and lowest variability of glucose values. In the next step, these three clusters were compared by means of C-peptide levels, the proportion of CGM measurements less than 70 mg/dl, between the interval of 70 mg/dl-180 mg/dl and higher than 180mg/dl by using non-parametric Kruskal-wallis test. Significant differences were found between the first and second clusters ($p=0.024$) and between the second and third (p=$0.065$).

Different than the previously mentioned studies, \citet{matabuena2023reproducibility} were concerned about the reproducibility of the results of CGM systems within non-diabetic patients. They applied the functional data analysis approach to test the interday reproducibility of glucose levels obtained from the subjects who participated to the clinical trial AEGIS \citep{gude2017glycemic} in Spain.
To do so, the intraclass correlation coefficient (ICC; see, for instance,  \citeauthor{matabuena2022estimating}, \citeyear{matabuena2022estimating}) for functional data
was computed using a two-way FANOVA multilevel model \citep{di2009multilevel}.
In models with repeated measures, the ICC quantifies the proportion of total variability explained by the subjectsâ€™ effect. High values of ICC are indicators of reproducibility.

\section{Other Types of Sensor Data}\label{sec:Other}
Different studies were related to data recorded by other types of sensors different than accelerometers and glucometers. These studies mainly focused on  blood pressure data and on electrodermal activity data which were recorded by ambulatory blood pressure monitor, spectrometry or armband. The functional approaches used in modelling these types of sensor data were exploratory FDA \citep{coffman2020challenges}, FPCA \citep{Mueller2011,Goldsmith2017,madden2018morning}, SoFR \citep{Tekwe2019}, FoFR \citep{Goldsmith2017}, functional logistic regression \citep{ries2023predicting}, and
curve registration \citep{Regis_et_al_2023_jrsssc}.

The first group of studies focused on electrodermal activity sensors. In \citet{Mueller2011} the main objective was the development and calibration of a noninvasive continuous-time glucose monitoring system in humans.
In this study a portable multisensor was attached to the upper arm of subjects.
This device includes, among other sensors, three different electrodes with deep, mid, and shallow skin penetration, respectively,
measuring the response of the skin and underlying tissue to an externally applied
electric field with frequencies in the range of 0.1â€“100 MHz.
Thus, each measurement provides three spectral functions (one per electrode) which are known to be related with BGL.
The analyzed sample consisted of eight patients (four males with Type I diabetes mellitus and four males with Type II diabetes mellitus).
Ten hours of data were recorded from four different visits of the patients.
The multisensor device took measurements every minute.
To ensure close monitoring of the BGL, it was measured from blood samples with a  standard technique every 15 minutes and then linearly interpolated to have BGL data every minute.
Three different functional data sets were considered, each containing one of the three spectral functions for each minute, visit and patient in the sample. FPCA was applied separately to each dataset, and the scores at the first two principal components were considered as features in a multiple linear regression to explain BGL, at which other measures provided by the multisensor were also included as potential explanatory variables.
The Akaike Information Criteria (AIC) was used to select the relevant subset of variables for blood glucose estimation.
This subset of relevant variables consisted exclusively of three out of six principal component scores.
The authors concluded that FPCA was useful for predicting BGL from the impedance spectra of human skin, and that it was feasible to monitor blood glucose variations with the presented noninvasive multisensor device.

\citet{Tekwe2019} aims to determine the importance of energy expenditure (EE) in obesity development on a sample of 374 children.
A SoFR model was used to assess the impact of EE  levels on BMI among elementary schoolâ€“aged children.
To be specific, the response variable in SoFR model was the logarithm of BMI and the functional regressor was the EE  per minute over a period of 30 hours (six hours at five weekdays in school).
The functional variable EE was not directly observable, but it was measured with error by the multisensor device SenseWear Arm Band (SWA, BodyMedia Inc., \url{www.bodymedia.com}) which includes sensors of heat flux, skin temperature and galvanic skin response.
Additionally, an accelerometer was integrated into SWA to count steps per minute.
The step counts were treated as a functional instrumental variable for the true EE.
The main contribution of this paper was a novel estimation method for the SoFR model, based on a function valued instrumental variable when the functional covariate is measured with error.

\citet{coffman2020challenges} considered electrodermal activity (EDA) data of adolescent mothers to understand their relationship with their children. Power Source Parenting Study data collected by \citet{Rajan2012} was used in the study. The objective of the study was to identify the peaks in mothers' EDA curves of mothers during different tasks, such as playing with their children, teaching their children a concept above their developmental level, and walking up and down stairs. A wrist worn iCalm biosensor device was used in record EDA measurements.
B-spline basis functions and local polynomial regression approaches were used in smoothing EDA profile of subjects in two different tasks (teaching and free play) at different time periods (baseline, 3-month follow-up, 6-month follow-up).
Exploratory FDA was used to identify peaks of EDA functions.

The second group of studies were using ambulatory blood pressure (ABP) monitoring  data.

\citet{Goldsmith2017} introduced a variational Bayes variable selection method for the functional linear concurrent model (FLCM, a special case of FoFR model) with sparse and irregularly observed functional data.
The proposed method was applied on Masked Hypertension Study \citep{schwartz2016clinic} where ABP measurements of 888 subjects were recorded every 28 minutes over a 24-hours monitoring period.
After each BP measurement, participants were asked to answer in an electronic diary questions about their situation, activities, affect, and social interactions.
Additionally, physical activity and sleep duration were monitored by actigraphy devices.
In the FLCM, up to 32 functional covariates (physical activity and entries in the electronic diary) were considered to predict the ABP levels as functional response.
The paper was complemented with the publicly available R package  {\tt vbvs.concurrent}.

In the study of \citet{madden2018morning} a sub-sample taken from Mitchelstown Ambulatory Blood Pressure Monitoring data \citep{kearney2013cohort} was used to predict morning blood pressure surge of subjects. ABP monitoring was done by using Meditech ABPM-05 (https://www.meditech.hu/) device that takes measurements every half an hour during 24 hours. Although the main objective of the study was to offer a multiple component random effects model alternative to a traditional model (where a cosine curve is fitted to periodic data), a FPCA based model was as well used for comparison.
Due to the sparse structure of the data, FPCA was done using the proposal of  \citet{yao_et_al:2005:sparseFDA}.

The third group of works focus on the analysis of heart rate (HR) signals.
The study of \citet{ries2023predicting} focused on application of functional logistic regression for predicting probability of fatigue from heart rate data of the subjects examined in the Wearables at the Grand Canyon for Health (WATCH) Study \citep{avina2017rim}.
In the study 47 participants who hiked the 20.7 mile Grand Canyon rim-to-rim trail were observed.
A total of 43 of them used a chest strap ECG device while four of them were using a wearable device on their wrist to record HR signals by Photoplethysmography (PPG), a light-based technology that measures peripheral blood volume variations.
Fatigue was defined as the indicator of whether a subject was left behind during the ascent in comparison to the group of runners with whom he or she was performing the descent.
The HR signals were smoothed by using B-spline functions. The accuracy of the model tested by using leave-one-out-cross validation and the correct classification rate of the fatigue using HR data was found to be $79\%$. Then, a functional logistic model with additional covariates of age and gender was used but these covariates were not found significant.

\citet{Regis_et_al_2023_jrsssc} focused in detection of HR anomalies from PPG signals.
The proposed subject-specific methodology began by segmenting the PPG signal into pulse-to-pulse intervals.
The pulse curves were then registered using a piece-wise linear method to align the running location of the pulse maxima.
The registered curves were expanded in a B-spline basis with 10 elements. The expansion coefficients, estimated by a state-space model using the Kalman filter, were used as pulse-specific shape parameters.
Finally, a support vector machine (SVM) is used to detect HR anomalies for each pulse curve, using shape parameters as inputs and ground truth as output (ECG for the same patients and periods were available, with every heart beat correctly classified by trained analysts).
After analyzing six-minute segments corresponding to 25 patients, the authors concluded that the shape parameters accurately extracted relevant information from the PPG (they reported
an average accuracy of $91.00\%$, with a standard deviation of $8.77\%$, in classifying pulse curves as normal or abnormal).

\section{Open Wearable Sensor Data Sets}\label{sec:Open}

 The increasingly common use of sensor data leads to the integration of wearable device data into existing open databases or the creation of new ones. One of those recent databases is the ``The Open Wearables Initiative" (OWEAR, \url{https://www.owear.org/}) which is an open platform founded in 2020. The aim of OWEAR is to facilitate access to the open source data and to the algorithms. It also includes wearable device data from the existing UCI Machine learning (\url{https://archive.ics.uci.edu}) and Physionet (\url{https://physionet.org}) websites, most of which is accelerometer-related data.

In the website there are various package links from R or Python that could be used in the analysis of wearable sensor data. Among the Python and R packages developed for processing sensor data and extracting features from them, the following worth to be mentioned.
In R software the packages
 {\tt pawacc} \citep{geraci2013pawacc},
 {\tt nhanesaccel} \citep{van2014nhanesaccel},
 {\tt PAactivPAL} \citep{zhang2015paactivpal},
 {\tt PASenseWear} \citep{zhang2016pasense},
 {\tt accelerometry} \citep{van2018package},
 {\tt GGIR} \citep{migueles2019ggir, GGIR}
 and
 {\tt PhysicalActivity} \citep{Choi_et_al:2021}
 were proposed to extract features from different types of accelerometers while the packages
 {\tt CGManalyzer} \citep{zhang2018cgmanalyzer},
 {\tt cgmanalysis} \citep{vigers2019cgmanalysis}
 and
 {\tt iglu} \citep{broll2021interpreting}
 were specifically developed for analyzing glucometer data.
 The R packages,
 {\tt CardiacProfileR} \citep{djordjevic2019cardiacprofiler}
 and
 {\tt mhealthtools} \citep{snyder2020mhealthtools}
 packages can be used to extract characteristics and activity profiles from sensors embedded in wearable devices such as cell phones and fitness trackers.
In Python the package
{\tt GaitPy} \citep{czech2019gaitpy}
is used to extract features specifically from accelerometer data,  the package
{\tt HeartPy} \citep{van2019heartpy} is used to analyze Photoplethysmography (PPG) signals associated with the cardiac activity that are recorded by smart devices while {\tt FLIRT} package \citep{foll2021flirt} provides various physiological features taken from different types of wearable device data such as electrodermal activity data, ECG data and accelerometer data.
In addition to these packages, \citet{biosensors:2022} created the R package {\tt biosensors.usc} to represent any type of biosensor data that come from different sources such as accelerometer, glucometer, ECG or smart phones as functional data by using distributional analysis techniques.

Additionally, in R software there are some packages including original data recorded by accelerometers. R package {\tt adeptdata} includes three different types of accelerometry data. First data set is based on outdoor running activity of a female adult recorded during 25 minutes by using ActiGraph GT9X sensors. The other two datasets include continuous walking activity and walking stride pattern records of 32 participants aged between 23 and 52 years.

Among the revised accelerometer  studies, $36\%$ of them were based on original data while the rest come from previous studies or clinical trials. The mostly used and easy accessible data sources are listed below:
\begin{itemize}
    \item National Health and Nutrition Examination Survey (NHANES) data set
    \newline (\url{https://www.cdc.gov/nchs/nhanes})
     \newline {Accelerometer: ActiGraph AM7164 uni-axial waist worn accelerometer}
    \vspace{0.3cm}
    \item Baltimore Longitudional Study of Aging (BLSA)
    \newline (\url{https://www.blsa.nih.gov})
    \newline{Accelerometer: Actiheart monitor uni-axial chest worn accelerometer}
    \vspace{0.3cm}
    \item UK Biobank Accelerometer Data
    \newline (\url{https://biobank.ndph.ox.ac.uk/showcase/})
    \newline {Acceleometer: Axivity AX3 tri-axial wrist worn accelerometer}
    \vspace{0.3cm}
    \item Human Activity Recognition Using Smartphones Data Set    \newline(\url{https://archive.ics.uci.edu/ml/datasets/human+activity+recognition+using+smartphones})
    \newline {Accelerometer: Samsung Galaxy S II with embedded  accelerometer}
\end{itemize}

Half of the revised glucometer related papers were coming from clinical trials while the other half were using original data sets. Two of these clinical trial data sets, used in the papers of \citet{matabuena2021glucodensities} and \citet{gecili2021functional}, are:

\begin{itemize}
    \item Juvenile Diabetes Research Foundation (JDRF) Continous Glucose Monitoring (CONCEPTT Randomized Control Trial)
    \newline (\url{https://public.jaeb.org/datasets/diabetes})

    \vspace{0.3cm}
    \item A Estrada Glycation and Inflammation Study (AEGIS) Data
    \newline (\url{www.clinicaltrials.gov}, code NCT01796184)
\end{itemize}
It is possible to reach these data sets from the websites of JAEB Center for Health Research (\url{https://public.jaeb.org/datasets/diabetes}) and US National Library of Medicine (\url{www.clinicaltrials.gov}). In the webpage of JAEB Center for Health Research 24 out of 31 listed data sets include CGM data.

As part of a research project, \citet{Awesome-CGM} created a publicly accessible database that consists of 12 different CGM datasets collected by different types of wearable devices to monitor BGL of subjects. Nine of the studies are related to Type I diabetes and three of them related to Type II diabetes (two of them on humans and one of them on pigs).

Finally, in the section of other types of sensor data, the proportion of the papers using the author's own data was $57\%$. The rest were coming from previous studies \citep{Rajan2012, kearney2013cohort, avina2017rim}.

Regarding libraries implementing FDA tools, it is worth mentioning the following.
The R packages {\tt fda} and {\tt fda.usc} are general purpose tools, covering from descriptive analysis to FPCA and standard functional regression models. In Python the package {\tt scikit-FDA} has similar objectives.
Registration is well covered in the R libraries {\tt registr} or {\tt fdasrvf}.
Packages focused on FPCA are {\tt fdapace} (for sparse functional data), {\tt mfaces} and {MFPCA} (these last two for multivariate FPCA).
Several R packages are devoted to functional regression models:
{\tt refund} and {\tt FDboost}
implement linear regression, generalized regression, and additive regression models for functional data;
{\tt fastFMM} fits functional mixed models using
fast univariate inference;
{\tt denseFLMM} and {\tt sparseFLMM}
estimate linear mixed models (for both dense and sparse functional responses, respectively) using FPCA;
{\tt multifamm} allows to fit multivariate functional additive mixed models based on multivariate FPCA.
Other R libraries on FDA can be found in the CRAN task view (\url{https://cran.r-project.org/web/views/FunctionalData.html}).

\section{Conclusion} \label{sec:Concl}
Wearable devices are increasingly a part of everyday digital medicine. FDA is a well-suited tool for analysing this kind of sensor data. Functional approaches allow researchers to work on high dimensional data without losing any information. Many classical approaches were extended to the functional case to be able to work with data as curves. FPCA is the main tool for dimensionality reduction and for revealing the most important modes of variation in functional data sets.  Multilevel FPCA can reveal patterns of variation at both individual and temporal levels that could not be analysed using traditional methods. Functional generalized linear models are used to model scalar or functional responses depending on different predictors. Functional generalized linear mixed models allow to consider repeated measurements and dependent observations. In general, functional models proposed in reviewed articles were good in prediction comparing to alternative statistical methods.

The literature on FDA-based analysis of wearable device data has been increasing enormously during the last years. It is possible to find many methodological contributions as well as relevant applications. In particular, the largest number of publications has been in the field of accelerometer data, followed by glucometer data and other type of sensor data (ECG data, ambulatory blood pressure data, skin sensor data, etc.). A total of 59 papers covering all of these data type have been included in this systematic review.
\backmatter


\section*{Declarations}


\begin{itemize}
\item Funding: This research was supported by the Spanish Research Agency (AEI) under projects PID2020-116294GB-I00 and PID2023-148158OB-I00, by AGAUR under grant 2021 SGR 00613, and by UPC under AGRUPS-2022 and AGRUPS-2023.
\item Competing interests: The authors have no competing interests to declare that are relevant to the content of this article.

\end{itemize}


\bibliography{references.bib}

\end{document}